\pdfoutput=1

\documentclass[reprint,aps,prl,showpacs,twocolumn]{revtex4-1}
\usepackage{mathptmx}
\usepackage{epsfig,amsopn}
\usepackage{graphicx}
\usepackage{amsmath,amssymb}
\usepackage{natbib,color}
\usepackage{braket}
\usepackage{float}
\usepackage{enumerate}

\allowdisplaybreaks[1]

\usepackage{adjustbox}
\usepackage[flushleft]{threeparttable}

\usepackage[framemethod=TikZ]{mdframed}
\newcounter{theo}
\renewcommand{\thetheo}{\arabic{theo}}

\def\bea{\begin{eqnarray}}
\def\eea{\end{eqnarray}}

\begin{document}

\title{Closed-form survival probabilities for biased random walks at arbitrary step number}

\author{{\normalsize{}Debendro Mookerjee}
{\normalsize{}}}

\author{{\normalsize{}Sarah Kostinski}
{\normalsize{}}}
\email{sk10775@nyu.edu}

\affiliation{\noindent \textit{Department of Physics, New York University, New York, USA}}

\date{\today}

\begin{abstract}

\noindent{We present a closed-form expression for the survival probability of a biased random walker to first reach a target site on a 1D lattice. The expression holds for any step number $N$ and is computationally faster than non-closed-form results in the literature. Because our result is exact even in the intermediate step number range, it serves as a tool to study convergence to the large $N$ limit. We also obtain a closed-form expression for the probability of last passage. In contrast to predictions of the large $N$ approximation, the new expression reveals a critical value of the bias beyond which the tail of the last-passage probability decays monotonically.}

\end{abstract}

\maketitle
Random walks appear in many fields of study, including financial economics \cite{Finance1, Finance2}, population genetics \cite{Genes, Genes2, Genes3, Genes4, Genes5}, neurobiology \cite{Redner}, semiconductor manufacturing \cite{Semiconductors, Semiconductors2, Semiconductors3}, and quantum physics \cite{Physics}. In these fields, random walks are used to model share prices, the statistical properties of genetic drifts, the firing of neurons, the diffusion of dopants and impurities in semiconductor fabrication, and the diffusion of molecules in fluids. Their ubiquity underscores their importance. 

The survival probability and first passage distribution are 
key properties of random walks that affect many stochastic processes. The survival probability is defined as the probability that the random walker has not reached a designated target site for a given duration. Its complement is the cumulative of the probability distribution of first passage times. To illustrate their application in neurobiology, random walks are used to model how stimulative and repressive inputs change the neuron's polarization above its reference level \cite{Neuron}. In that case, the distribution of times between neural firings represents a first passage problem, since a neuron emits an electrical signal when its polarization level first exceeds a threshold level that is greater than its initial reference level~\cite{Redner}. First passage processes also appear in random walk models for diffusion-controlled reactions \cite{Redner}: In these reactions, diffusing reactant particles are 
converted into product particles when they meet for the first time. 
Other applications of first passage processes can be found in, e.g., Ref.~\cite{Applications}.

Here we consider the survival probability of a biased random walker on a discrete 1D lattice to first reach a target site other than its origin. Surprisingly, exact expressions in closed form for such probabilities at arbitrary step number are lacking. Rather, the primary focus in the literature has been on the limit of large step number, or long times~\cite{MajumdarWedge}. Below we present a path enumeration method to provide new closed-form expressions for the survival probability at any step number, including the elusive intermediate range.

\begin{figure}[b!]
\begin{centering}
\includegraphics[width=0.49\textwidth]{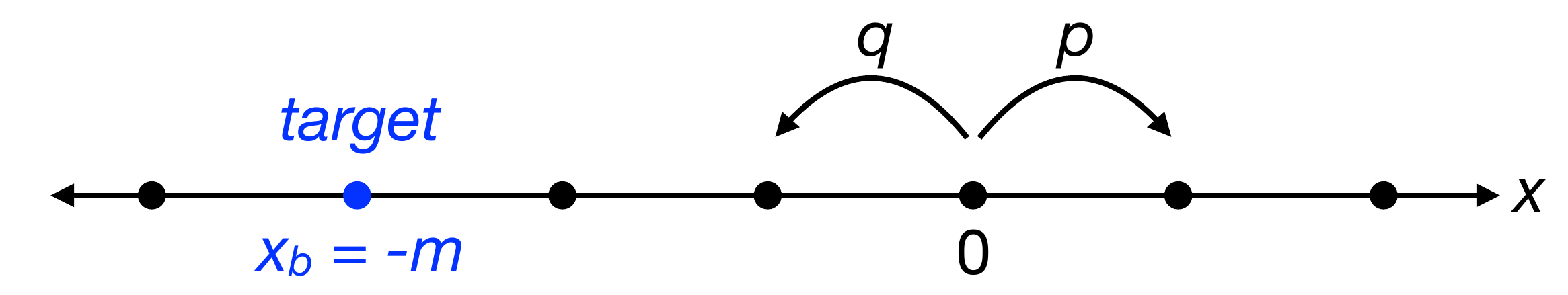}
\end{centering}
\vspace{-2mm}
\caption{A random walk on a discrete one-dimensional lattice, with probability $p$ of stepping to the right, and probability $q=1-p$ of stepping to the left. Here we consider biased random walks, i.e. $p \neq q$, which start at the origin. We focus on the survival probability to remain to the right of a target $x_b=-m$ (marked in blue), where $m$ is a positive integer.}
\label{Lattice}
\end{figure}

Obtaining a closed-form expression of the survival probability that holds for arbitrary step number is not always tractable by the usual generating function approach. To quote Feller from his text on probability theory~\cite{Feller}: ``Given a generating function $P(s) = \sum p_k s^k$ the coefficients $p_k$ can be found by differentiations from the obvious formula $p_k = P^{(k)}(0)/k!$.  In practice it may be impossible to obtain explicit expressions and, anyhow, such expressions are frequently so complicated that reasonable approximations are preferable.'' His remark implies several disadvantages with the generating function approach: differentiating $k$ times becomes cumbersome, especially if the generating function is complicated, and typically one must make approximations based on expressions in non-closed-form. Here we aim to show that an accessible closed-form expression for the survival probability at arbitrary step number can be obtained using a combinatorial path enumeration approach. 

\begin{figure}[t!]
\begin{centering}
\includegraphics[width=0.49\textwidth]{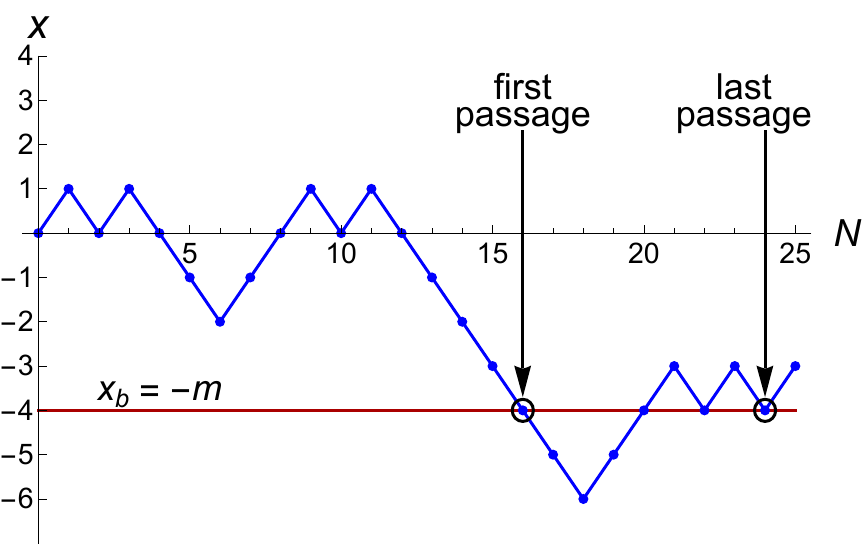}
\end{centering}
\vspace{-5mm}
\caption{Sample trajectory of a random walker on a discrete one-dimensional lattice, with target position $x_b = -4$ demarcated by the solid red horizontal line. Step number is denoted as $N$ on the abscissa, and position $x$ (net displacement from the origin) is given by the ordinate. First passage occurs when the random walker reaches the target $x_b = -4$ for the first time.}
\label{Traj}
\end{figure}

To derive such an expression, consider a random walker on the real number line whose starting position $x_0$ is $x_0 = 0$. The random walker jumps to the right or left to a 
neighboring position with probabilities $p$ and $q$, respectively (see Figure~\ref{Lattice}). Consider a target position $x_b = -m < 0$, where $m$ is a positive integer. We will consider the survival probability to remain to the right of $x_b$ up to step $N$. This scenario is equivalent to a random walker starting at integer $+m$ and surviving until reaching the position $x = 0$. Figure \ref{Traj} shows a sample trajectory for the case $m = 4$.

Let us define $R(N)$ as the probability that the random walker has remained to the right of $x_b = - m$ for $N$ steps. $R(N)$ is then simply the sum of the probabilities of all possible surviving paths. For a fixed number of total steps $N$, a survival path can consist of $n$ and $k=N-n$ steps to the right and left, respectively. The probability of such a survival path is $p^nq^k = p^{N-k}q^k$. By representing such a path as a sequence of $n$ $(+1)$'s and $k$ $(-1)$'s, we see that a survival path requires that its sequence's running sum remains greater than or equal to $1- m$. The number of such ``survival'' sequences is given by the $m$th Catalan trapezoid $C_m(n,k)$ \cite{Cat}:
\begin{align}
    C_m(n,k) = \begin{cases}
        \binom{n+k}{k} & {0 \le k < m}\\
        \binom{n+k}{k} - \binom{n+k}{k-m} & {m \le k \le n + m - 1}\\
        0 & k > n + m - 1
    \end{cases}
    \label{CatalanTrapezoid}
\end{align}
Consequently, the probability of a survival path remaining to the right of $x_b = -m$ with $n$ steps to the right and $k=N-n$ steps to the left is $p^{N-k}q^k C_m(N-k,k)$. Therefore, the survival probability $R(N)$ becomes: 
\begin{align}\label{Catalan}
    R(N) &= \sum_{k} p^{N-k}q^k C_m(N-k,k) \, .
\end{align}
The value of the survival probability thus depends on how $m$ compares to $N$.

Let us first consider the case when $m > N$. Only the $0 \le k < m$ case for the Catalan trapezoid applies and hence $C_m(n,k) = \binom{n+k = N}{k}$. 
Using the binomial theorem, $R(N)$ simplifies to:
\begin{align}
    R(N) = \sum_{k = 0}^N p^{N-k}q^k \binom{N}{k} = (p + q)^N = 1 \, .
\end{align}
This is indeed expected: a random walker will never reach the target position $x_b = -m$ with an insufficient number of steps to reach the target position. 

We now consider the case when $m = N$. Both the $0 \le k < m$ and $k = m$ cases of the Catalan trapezoid (see Eq.~\ref{CatalanTrapezoid}) apply in this case. Using the binomial theorem once more, we see that 
\begin{align}
    R(N) &= \sum_{k} p^{N-k}q^k C_m(N-k,k) \nonumber \\
    &= \sum_{k = 0}^{m-1} p^{N-k}q^k \binom{N}{k} + q^N\bigg[\binom{N}{N} - \binom{N}{0}\bigg]\nonumber\\
    &= \sum_{k = 0}^N p^{N-k}q^k \binom{N}{k} - q^N \nonumber \\
    &= 1 - q^N
\end{align}
This result is again expected: If the random walker always jumps to the left for $N = m$ steps, he will reach the 
target $x_b = -m$. There is only one such path, composed solely of steps to the left, which occurs with probability $q^N$. Since the random walker either reaches the target $x_b$ or survives, $q^N + R(N) = 1$. 

The interesting case occurs when 
$m < N$. Both the $0 \le k < m$ and $m \le k \le n + m - 1$ cases for the Catalan trapezoid apply and the survival probability becomes:
\begin{align}
    R(N) &= \sum_{k = 0}^{m-1}p^{N-k}q^k \binom{N}{k} + \sum_{k = m}^{n+m-1} p^{N-k}q^k \bigg[\binom{N}{k} - \binom{N}{k - m}\bigg] \nonumber\\
    &= \sum_{k = 0}^{n+m-1}p^{N-k}q^k \binom{N}{k} - \sum_{k = m}^{n+m-1} p^{N-k}q^k\binom{N}{k - m} \, .
\end{align}
The upper summation limit $k \le n + m - 1$ can be expressed in terms of $N = n + k$; that is, $k \le N - k + m -1$. This condition simplifies to $2k \le N + m - 1$, or $k \le \frac{N+m-1}{2}$. Since $k$ must be an integer, we round down this upper limit using the floor function. The survival probability thus becomes:
\begin{align}
    R(N) = \sum_{k = 0}^{\lfloor \frac{N+m-1}{2} \rfloor}p^{N-k}q^k \binom{N}{k} - \sum_{k = m}^{\lfloor \frac{N+m-1}{2} \rfloor}p^{N-k}q^k \binom{N}{k - m} \, .
\end{align}
Defining $\gamma \equiv \frac{q}{p}$, the survival probability can be expressed as
\begin{align}\label{Sum}
    R(N) &= p^N\sum_{\ell=0}^{\lfloor \frac{N+m-1}{2} \rfloor} \gamma^\ell \binom{N}{\ell} - p^{N-m}q^m\sum_{\ell = 0}^{\lfloor \frac{N+m-1}{2} \rfloor - m} \gamma^\ell \binom{N}{\ell}
\end{align}
\begin{figure*}[t!]
\begin{centering}

\includegraphics[width=\textwidth]{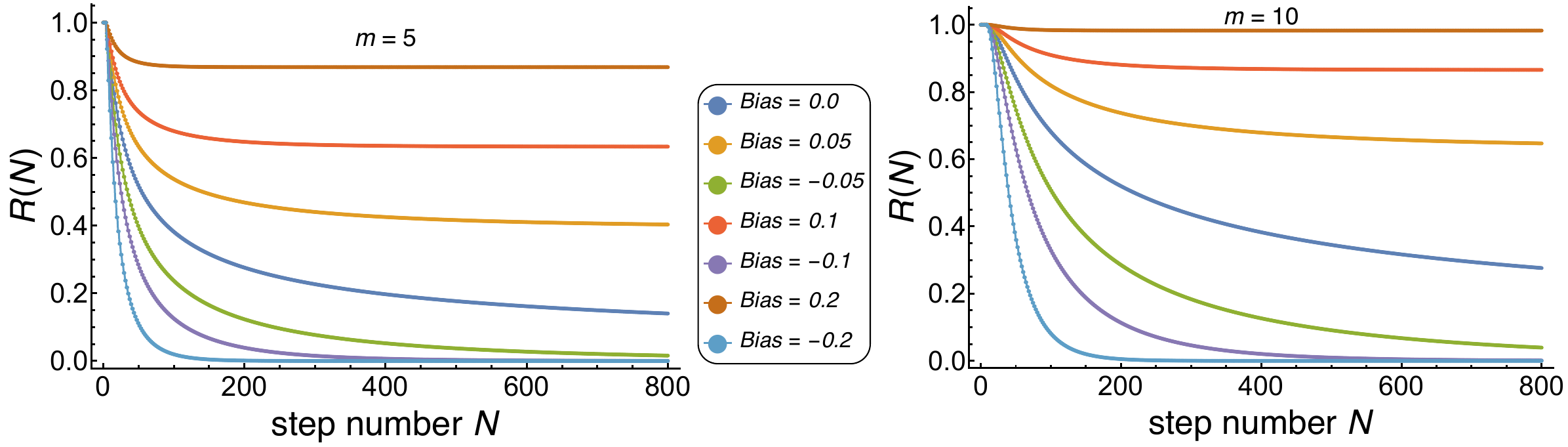}
\end{centering}
\vspace{-5mm}
\caption{Plots of the 
survival probability $R(N)$ as a function of step number $N$ for biases $-0.2 \le B \le 0.2$. The left panel examines the case $m = 5$, while the right panel is plotted for $m = 10.$ A non-zero saturation limit exists for positively biased random walks and this limit increases as $m$ grows. Though not easily visible in these plots, survival probabilities at adjacent step numbers are equal when $N \ge m$. }
\label{Sur}
\end{figure*}
We find the summation in Eq.~\ref{Sum} can be expressed in terms of Gauss hypergeometric functions~\cite{Hypergeometric}. The arguments of the hypergeometric functions depend on the parity of the sum $N + m$. For the sake of 
brevity, the full derivation is provided in the Supplemental Material \cite{SupplMat} and we present here the end result. When $N + m$ is even, we find the survival probability to be 
\begin{widetext}
\begin{align}\label{even}
    R_{e}(N) &= 1 - \gamma^m + p^N \gamma^\frac{N+m}{2} \binom{N}{\!\frac{N-m}{2}\!}\bigg[ 1 - {}_2F_1 \bigg(\! \!1; \frac{m-N}{2}; \frac{2+N+m}{2}; -\gamma\bigg)  \bigg] + p^N\gamma^\frac{N+m+2}{2} \binom{N}{\frac{2+N-m}{2}} {}_2F_1 \bigg(\!\!1; \frac{2 - N - m}{2}; \frac{4+N-m}{2}; -\gamma\bigg)
\end{align}
\end{widetext}
where the subscript $e$ in $R_{e}$ denotes that $N+m$ is even. When $N + m$ is odd, we find the survival probability to be 
\begin{widetext}
\begin{align}\label{odd}
    R_{o}(N) &= 1 - \gamma^m + p^N\gamma^\frac{N+m+1}{2}\binom{N}{\frac{N-m+1}{2}}\bigg({}_2F_1 \bigg(1; \frac{1 - N - m}{2}; \frac{3+N-m}{2}; -\gamma\bigg) + \frac{1}{\gamma}\bigg[ 1 - {}_2F_1 \bigg(1; \frac{m-N-1}{2}; \frac{N+m+1}{2}; -\gamma\bigg)  \bigg]\bigg)
\end{align}
\end{widetext}
where the subscript $o$ in $R_{o}$ denotes that $N + m$ is odd. Eqs.~\ref{Catalan}, \ref{even}, and \ref{odd} reveal a relationship between hypergeometric functions and Catalan trapezoids. The latter two equations are exact and in closed form, and hold for any step number $N$ and any $p$ and $q$. We were not able to find equivalent closed-form expressions for the survival probability in the literature. The expressions for $R(N)$ given above in terms of $p$ and $q$ can also be rewritten in terms of just the bias $B = p - q$. In Eqs.~\ref{even} and \ref{odd}, this simply requires substituting $p = (1 + B)/2$ and $\gamma = q / p = (1-B)/(1+B)$. We also note that in the case the target lies to the right of the origin, i.e. $x_b = +m$, the survival probability is defined as remaining to the left of $x_b$ for the duration of the walk. That probability, which we call $L(N)$, is equivalent to the expressions derived above for $R(N)$, but with the following substitutions: $\gamma = p/q$ instead of $q/p$, and $p$ in the prefactors multiplying the hypergeometric functions should be replaced by $q$ (see Eqs.~\ref{even} and \ref{odd}).

Plots of the 
survival probability $R(N)$, such as those provided in Figure~\ref{Sur}, reveal that for positive bias ($p > q$), the survival probabilities saturate to nonzero values. We analytically determined these values to be
\begin{equation}
1 - \gamma^m = 1 - \bigg(\frac{1-B}{1+B}\bigg)^m
\end{equation}
using Stirling's approximation. A full derivation is provided in the Supplemental Material \cite{SupplMat}. For a fixed positive bias $B$, we see that the saturation limit grows as the distance $m$ between the target 
and the walk's origin increases. This property is shown in Figure~\ref{Sur}. The same plots also show that for non-positive bias, i.e. $p \leq q$, the survival probability goes to $0$ in the limit of large step number. This reveals that in the limit of infinite step number, there exists no survival paths for negatively biased or unbiased walks, whereas there always exists some survival paths for positively biased walks.

Though not easily visible in Figure~\ref{Sur}, 
survival probabilities at adjacent step numbers are equal when $N \ge m$. When $m$ is odd, arriving at the target $x_b=-m$ occurs when the net displacement from the origin is odd. This can only happen when the random walker has jumped an odd number of steps. Consequently, the survival probability at odd step number $i$ is equal to the survival probability at the next consecutive even step number $i + 1$. Similarly, when $m$ is even, arrival at $x_b=-m$ occurs when the net displacement is even, which can only happen when the step number $N$ is even. As a result, the survival probability at even step number $i$ is equal to the survival probability at the next consecutive odd step number $i + 1$.  

\begin{figure}[t!]
\begin{centering}
\includegraphics[width=0.49\textwidth]{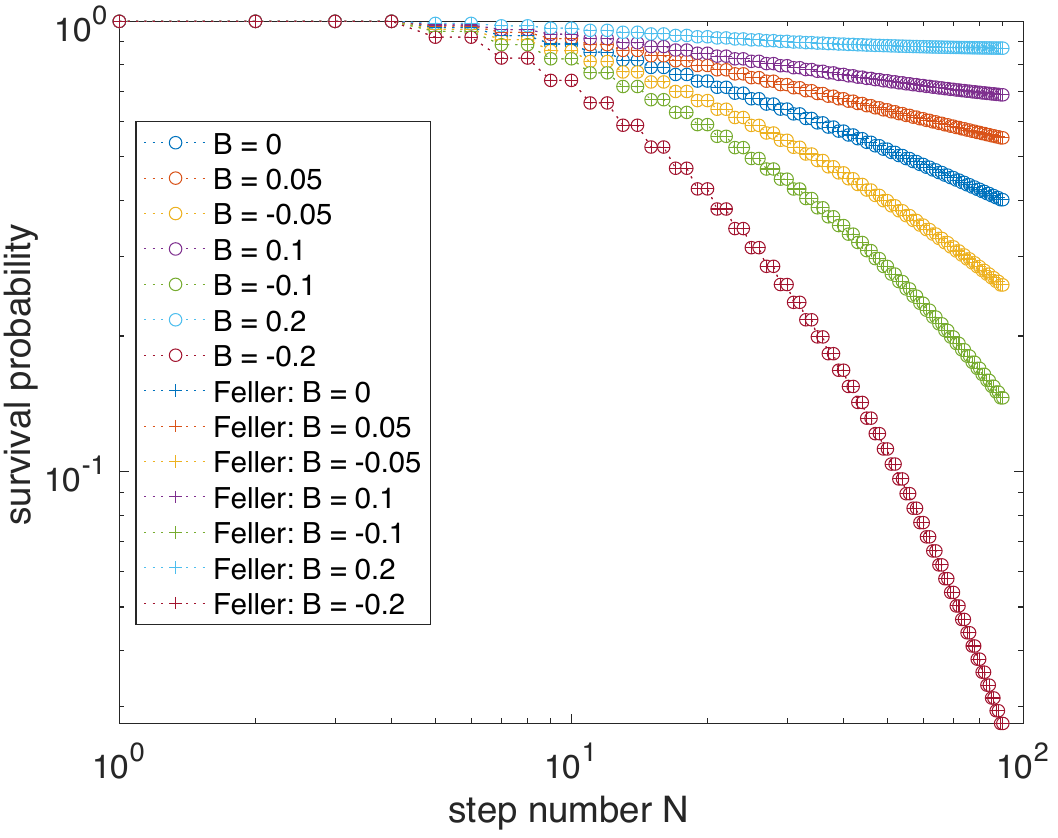}
\end{centering}
\vspace{-7mm}
\caption{Comparison between our closed-form expressions for the first passage survival probability $R(N)$ and Feller's non-closed expression for $m = 5$. The agreement 
illustrates the relationship between Catalan trapezoids, Gauss hypergeometric functions, and trigonometric integrals. We compared our results with Feller's for $N \le 90$. Beyond $N = 90$, Feller's formula requires long computation times.}
\label{Feller}
\end{figure}

We 
verify our results in Eqns.~\ref{even} and \ref{odd} through comparison to non-closed-form results in Feller's text~\cite{Feller}.  
These results are provided in the context of a gambler's ruin against an infinitely rich adversary (representing the infinite lattice), where the gambler's ruin probability at the $n$th trial is denoted by $u_{z,n}$. Here $z$ is the gambler's initial capital, which is analogous to the starting position of a random walker. To map this scenario to ours, the gambler's ruin problem is that of a random walker starting at $z = + m$, who survives up to step $n$ if he has not yet reached the origin (i.e. losing all of his capital). This is equivalent to our case of a random walker starting at the origin $x=0$ and not reaching a target $x_b = -m$ up to step $N$. The ruin probability at step $n$ given by Feller~\cite{Feller} is
\begin{align}
    u_{z,n} &= 2^n p^\frac{n-z}{2} q^\frac{n+z}{2}\int_0^1 \cos^{n-1}{\pi x} \cdot \sin{\pi x} \cdot \sin{\pi x z} \cdot dx \, .
\end{align}
Recall that we defined $R(N)$ as the survival probability at the $N$th step, and we now define its complement $\Tilde{R}(N)$ as the probability of being ruined by the $N$th step: $R(N) + \Tilde{R}(N) = 1$. The probability that the gambler is ruined at some step up to (and including step number) $N$ is thus 
\begin{equation}
\Tilde{R}(N) = \sum_{n = 1}^N u_{m,n} \, .
\end{equation}
Consequently, the survival probability based on Feller's non-closed expression for the ruin probability is:
\begin{align}\label{trig}
    R(N) = 1 \! - \! \sum_{n = 1}^N 2^n p^\frac{n-m}{2} q^\frac{n+m}{2} \! \! \! \int_0^1 \! \! \! \cos^{n-1}{\pi x}  \cdot \sin{\pi x} \cdot \sin{m\pi x } \, \, dx \, .
\end{align}
We numerically compare this result (Eq.~\ref{trig}) with our closed-form expressions (Eqs.~\ref{even}-\ref{odd}), which show excellent agreement. A sample 
plot is shown in Figure~\ref{Feller} for the case $m = 5$ and biases ranging from $B=-0.20$ to $0.20$. 
The agreement illustrates the close connections between Catalan trapezoids, hypergeometric functions, and trigonometric integrals.

Additionally, the observation that the survival probability at odd step number $i$ is equal to the survival probability at the next consecutive even step number $i + 1$ 
is more apparent from this figure. 

In Figure~\ref{Feller} we have compared the results only 
up to step number $N = 90$ due to the long computational times required to evaluate Feller's formula (Eq.~\ref{trig}) beyond this range. We note, however, that the computational times for Eqs.~\ref{even} and \ref{odd} were $\sim 10^{-5}$ to $10^{-6}$ times shorter and numerically these expressions could be computed for much larger step numbers such as $N=900$. 

This reveals a benefit of using exact closed-form expressions in lieu of their non-closed counterparts, and highlights an advantage 
our combinatorial approach has over the commonly-used generating function method. 

\begin{figure}[b!]
\begin{centering}
\includegraphics[width=0.5\textwidth]{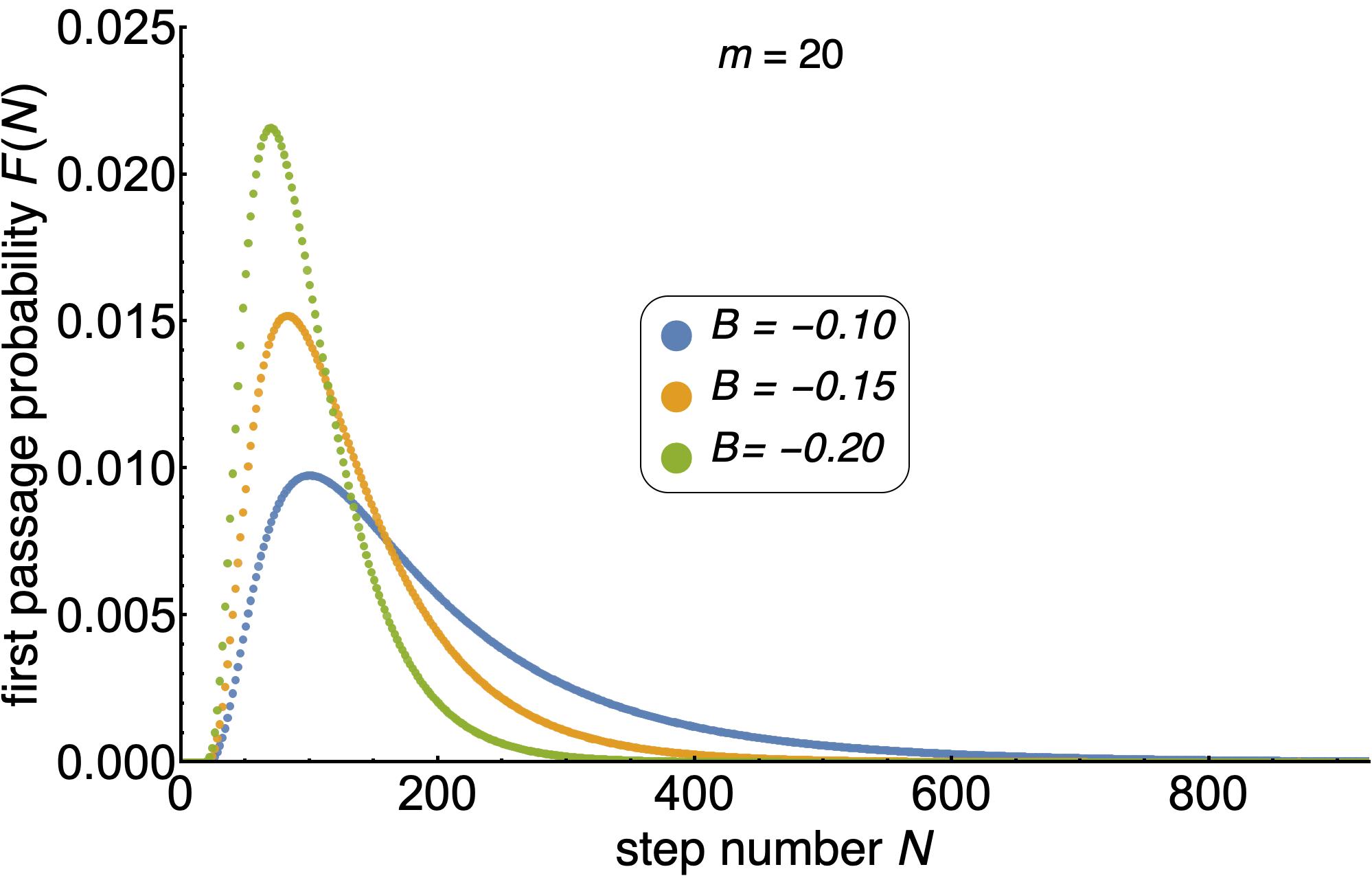}
\includegraphics[width=0.5\textwidth]{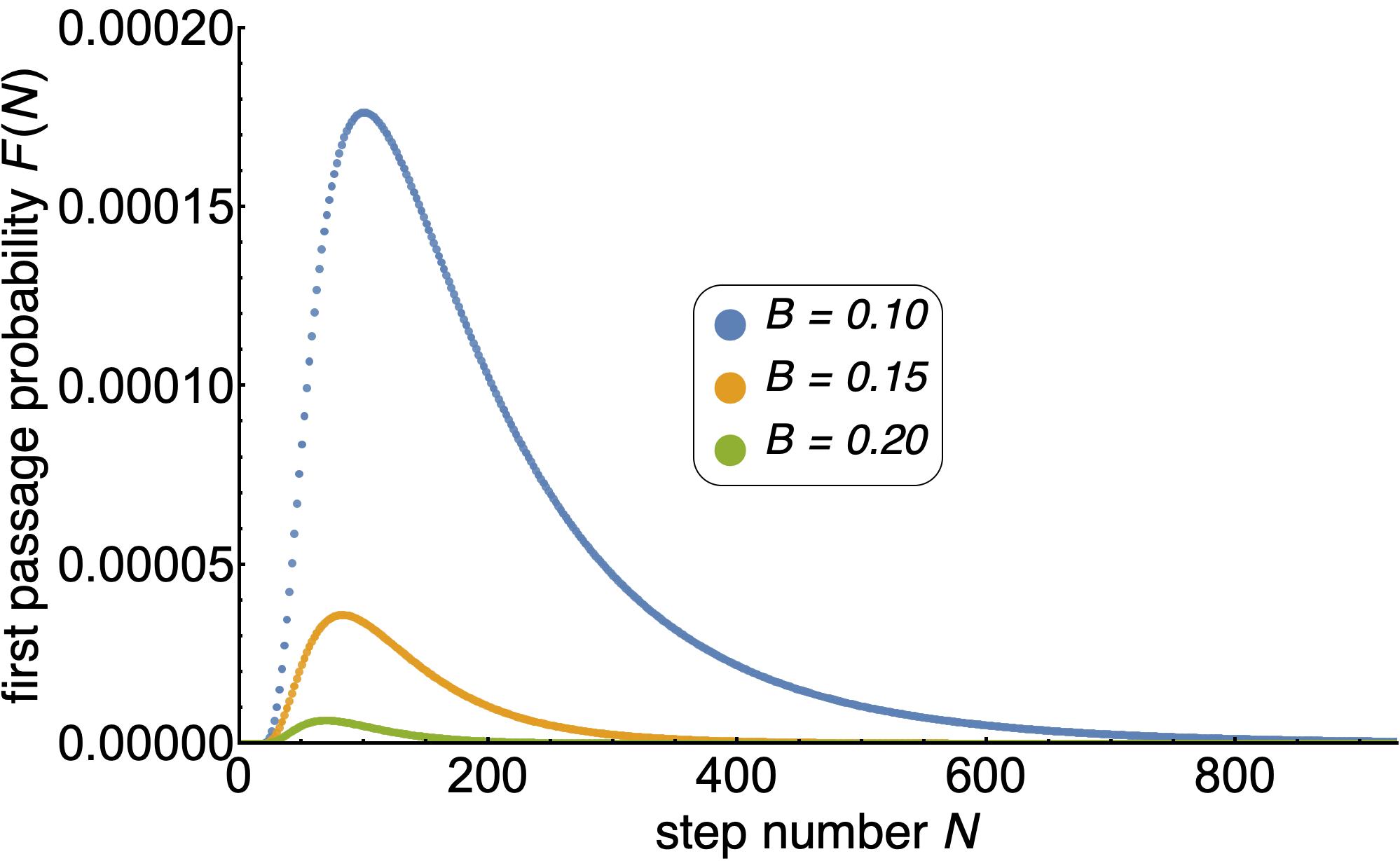}
\end{centering}
\vspace{-5mm}
\caption{Displayed here are the first passage probabilities $F(N)$ for the case $m=20$. We plot values for even-valued $N$, as first passage can only occur when $N+m$ is even ($F(N) = 0$ otherwise). The top panel shows results for negative values of the bias ranging from $B=-0.1$ to $B=-0.20$, while the bottom panel shows results for positive biases of the same magnitudes.
}
\label{FPTD}
\end{figure}
We now turn our attention to first passage probabilities. Note that from the survival probability $R(N)$, one can easily obtain the probability $F(N)$ of first hitting the target $x_b$ at step number $N$. Since $\sum_{j=1}^{N} F(j)$ is the probability of ruin at step $N$ and the complement of the survival probability, it follows that
\begin{equation}
R(N) = 1 - \sum_{j=1}^{N} F(j) \, .
\end{equation}
Consequently, we determine $F(N)$ from the difference between survival probabilities at consecutive step numbers, denoted here by $j$ and $j+1$: 
\begin{equation}
    F(j+1) = -(R(j +1) - R(j)).
    \label{eq:Sdiff}
\end{equation} 
Plots of $F(N)$ are shown in Figure~\ref{FPTD}. As negative biases increase in magnitude, the peak value of $F(N)$ shifts to the left. This is intuitive: The stronger the bias towards the target, the more likely the random walker is to reach $x_b = -m$ at an earlier step. 
Similarly, as positive biases increase in magnitude, the peak value of $F(N)$ also shifts to the left. Both cases 
illustrate that reaching the target for the first time is more likely to occur earlier in the walk than at later steps. However, the difference in the peak values for the two cases reveals that negatively biased walks are much more likely to reach the target than positively biased ones. In the limit of infinite step number, while $\sum_{N} F(N) = 1$ for negatively biased walks, for positively biased walks this sum remains less than 1. Thus, for $N \rightarrow \infty$, there exist no survival paths for negatively biased random walks, which tend to drift towards the target, while there are many survival paths for positively biased walks, which tend to drift away from the target.

We now consider the probability of last passage. We consider the probability for the random walker to return to the boundary of ruin $x_b = -m$ for the last time at step number $n$ for a walk length of $N$ steps. We call this probability the PLP for short. The PLP is a product of two probabilities, i.e. the probability to be at $x_b$ at step number $n$ and the probability to survive for the remainder of the walk:
\begin{align*}
    PLP &= \text{probability to be at $x_b$ at step number $n$}\\
    & \hspace{3mm} \times \text{probability of surviving $N - n$ steps thereafter}
\end{align*}
For $n < m$, the PLP is simply $0$ because there are too few steps to reach the boundary $x_b$. For $n > m,$ the PLP is
\begin{align}\label{PLPEq}
    PLP &= p^{\frac{n - m}{2}} q^{\frac{n + m}{2}} \binom{n}{\frac{n - m}{2}} \times S(N - n) 
\end{align}
where $S(N-n)$ is the survival probability to remain on the right or left side of $x_b$. This survival probability is different from $R(N)$, because the latter concerns a first passage problem (i.e. the random walk begins at an initial position different from the target), whereas the survival probability in Eq.~\ref{PLPEq} concerns recurrence (beginning and ending at $x_b$). 
We note that the exponents of $p$ and $q$ must be nonnegative integers because they represent the number of steps taken to the right, $n_R$, and left, $n_L$, respectively. Consequently, when $(n \pm m)/2$ are half-integers the PLP is zero because no such path can ever return to the boundary $x_b$. More details are provided in the Supplemental Material \cite{SupplMat}. 

The survival probability $S(N)$ appearing in Eq.~\ref{PLPEq} can be expressed in closed form for odd and even $N$ as~\cite{EPL}: 
\begin{align}\label{SOdd}
    S_{\text{odd}}(N,B) &= B + \frac{(1 - B^2)^\frac{N}{2}}{2^{N-1}} \binom{N-1}{\frac{N-1}{2}} \sqrt{\frac{1 - B}{1 + B}}  \bigg[1 - \nonumber \\
    &\frac{2B}{1 + B} \, \frac{N-1}{N+1} \, {}_2F_1\bigg(1;- \frac{N}{2}+\frac{3}{2} ;\frac{N}{2} + \frac{3}{2};\frac{B - 1}{B + 1}\bigg)\bigg]
\end{align}
\begin{align}\label{SEven}
    S_\text{even} (N,B) &= B +  \frac{(1 - B^2)^{\frac{N}{2}}}{2^N}\binom{N}{\frac{N}{2}}\bigg[\frac{1-B}{1 + B}  - 2B \, \frac{1 - B}{(1 + B)^2} \nonumber\\
    &\times \frac{N-2}{N+2}  \, {}_2F_1\bigg(1; -\frac{N}{2}+2;\frac{N}{2}+2;\frac{B-1}{B+1}\bigg)\bigg] \, .
\end{align}
\begin{figure}[t!]
\begin{centering}
\includegraphics[width=0.49\textwidth]{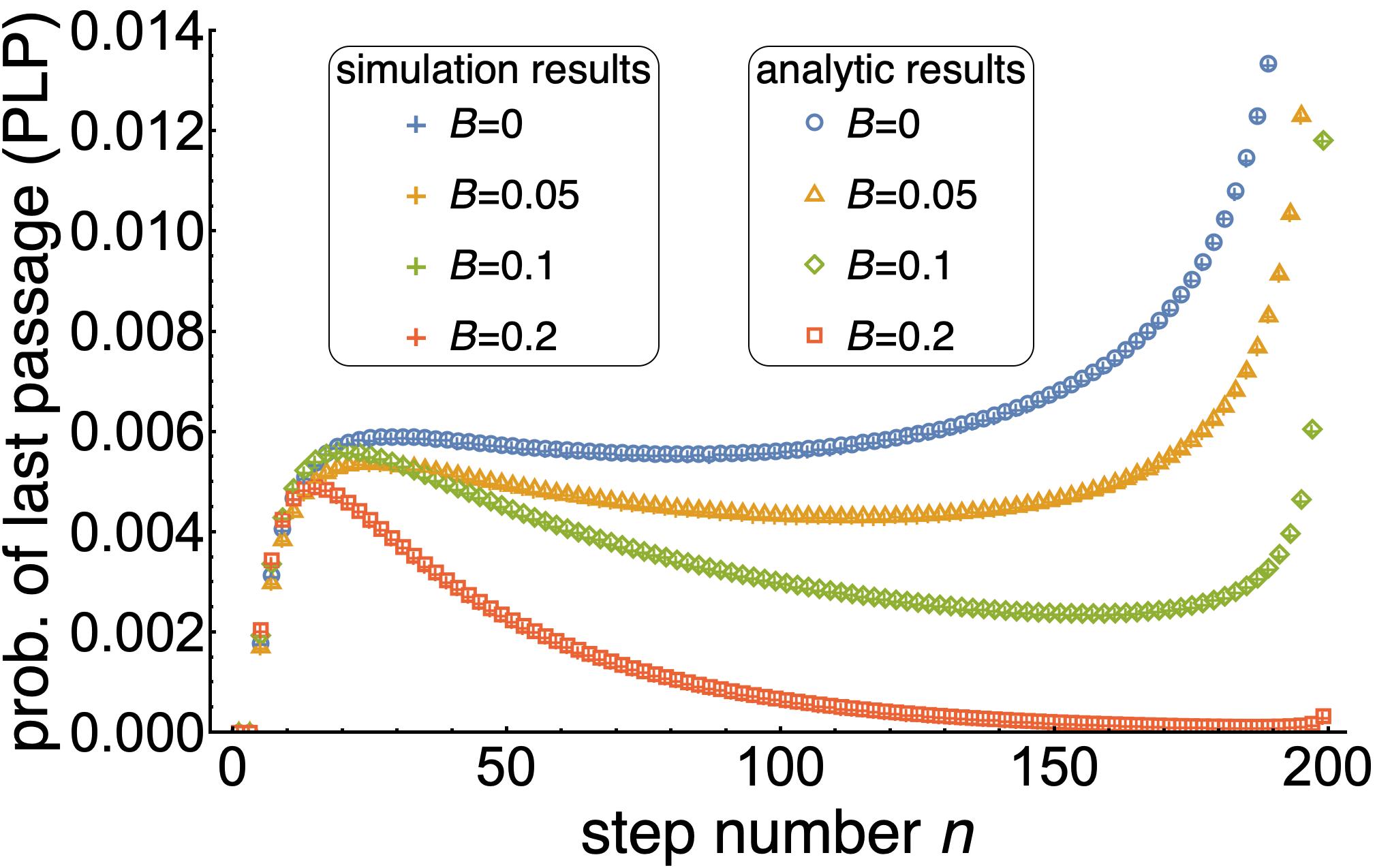}
\includegraphics[width=0.49\textwidth]{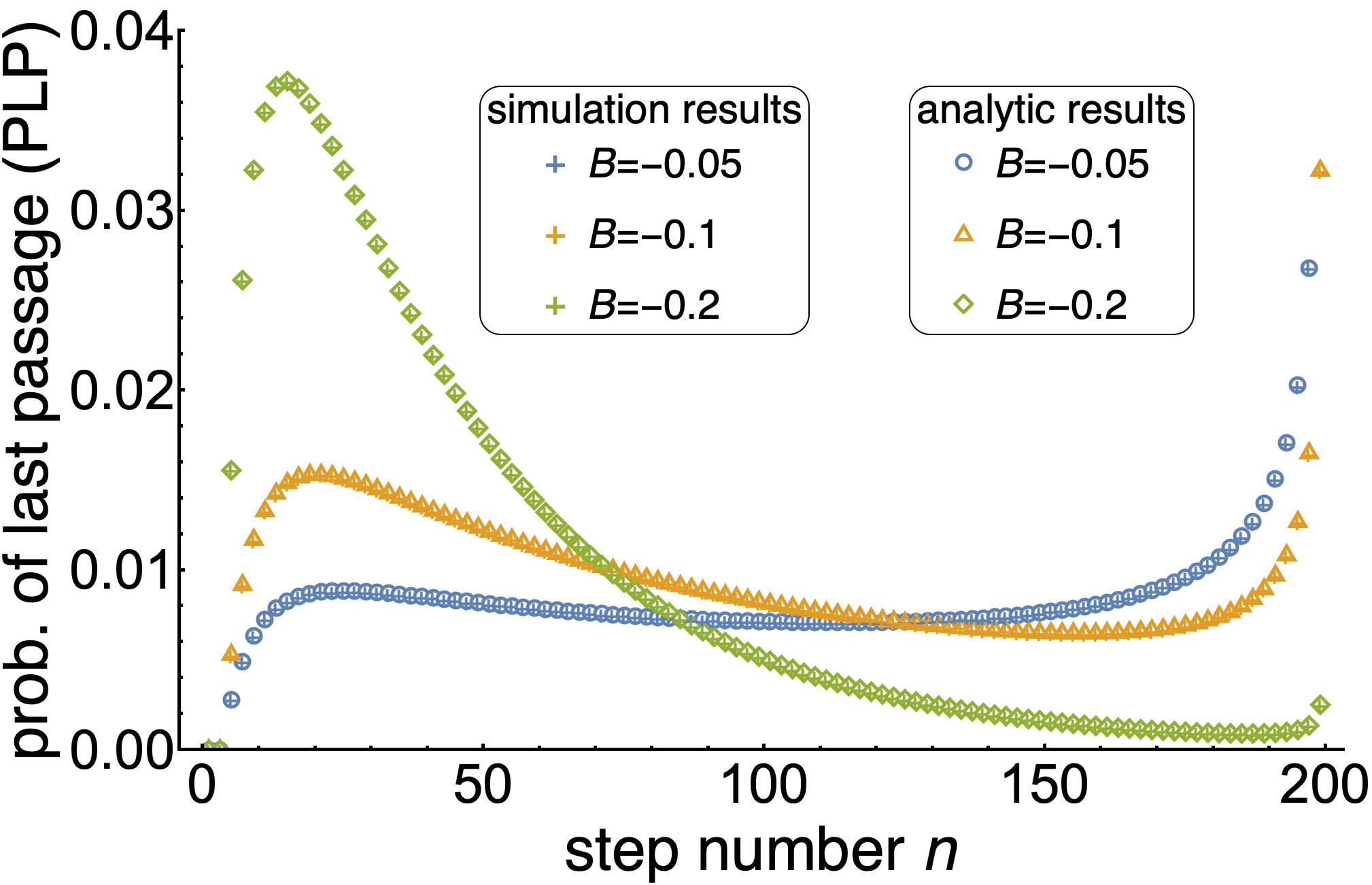}
\end{centering}
\vspace{-4mm}
\caption{The probability of last passage (PLP) for a random walk with target $m=5$ and total walk length $N=200$, plotted here for several positive biases (top panel) and negative biases (bottom panel). For visual clarity, we show only non-zero values of the PLP: When $m$ is odd, last passage occurs at odd step numbers and thus the PLP is zero for even step numbers. Note that as the magnitude of the bias is increased, the right peak of the PLP decays until it vanishes above a critical bias.}
\label{PLP}
\end{figure}
With these expressions for the survival probabilities $S_\text{odd}$ and $S_\text{even}$, we obtain closed-form expressions for the PLP. When plotted as a function of the step number $n$ (see Figure~\ref{PLP}), the PLP typically increases to a local maximum and decreases until near the end of the walk, where it again increases. Plots of the PLP for a total walk length of $N = 200$ steps with target $m=5$ and various biases are shown in Figure~\ref{PLP}. Note that we plot solely non-zero values of the PLP: The walker can only reach the target site at step $n$ if $n + m$ is an even number, otherwise the PLP is zero. The plots show that the rightmost peak of the PLP near the end of the walk decreases as the magnitude of the bias increases. Beyond a critical bias the rightmost peak disappears altogether. 

To quantify the critical bias, we study the case of a monotonically decreasing tail of the PLP. 
Specifically, we examine the difference between the PLP's penultimate and final (non-zero) values, e.g. when $n=N-2$ and $n=N$ in the case that $N+m$ is even. The sign of this difference determines whether the right tail of the PLP monotonically decreases. Such a condition yields the inequalities $PLP(N - 2) - PLP(N) > 0$ if $N+m$ is even, or $PLP(N - 3) - PLP(N-1) > 0$ if $N+m$ is odd. 
Using Eqs.~\ref{PLPEq}, \ref{SOdd}, and \ref{SEven}, we find that these inequalities reduce to:
\begin{align}
         |B|& > \sqrt{\frac{N^2 - 2N + m^2}{3N^2 - 2N - m^2}} & \text{for even $N+m$} \label{BcEvenSum} \\
         |B|& > \sqrt{\frac{N^2 - 4N + 3 + m^2}{3N^2 - 8N + 5 - m^2}} & \text{for odd $N+m$} \, \label{BcOddSum}.
\end{align}
The full derivation, which we omit here for brevity, can be found in the Supplemental Material~\cite{SupplMat}. 

\begin{figure}[t!]
\begin{centering}
\includegraphics[width=0.5\textwidth]{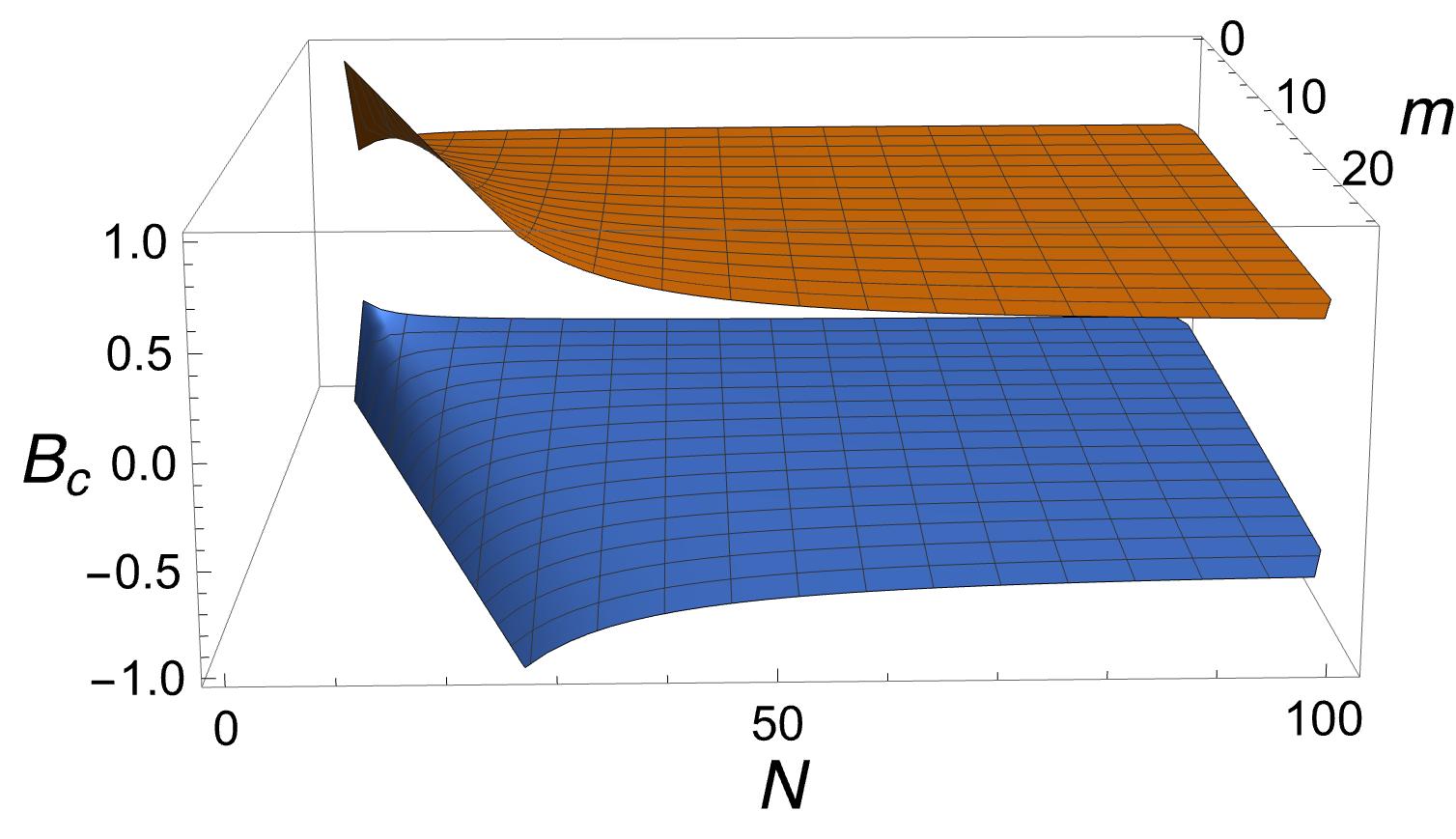}
\includegraphics[width=0.47\textwidth]{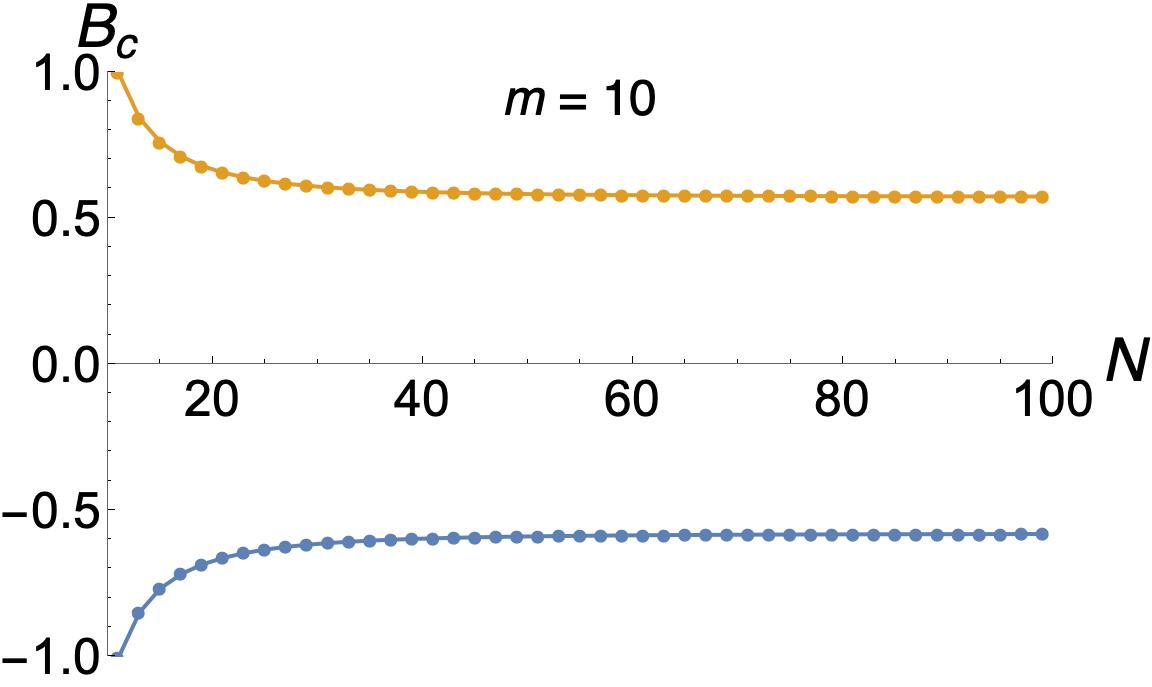}
\end{centering}
\vspace{-3mm}
\caption{The top panel illustrates the dependence of the critical bias (Eq.~\ref{BcOddSum}) on $m$ and $N$, where $N>m$ and the bias can be positive or negative (represented by the top and bottom surfaces, respectively). Regions above the top (orange) surface and below the bottom (blue) surface represent biases for which the tail of the PLP monotonically decreases. The bottom panel displays a cross-section of the two surfaces for $m=10$. The critical bias approaches $1/\sqrt{3}$ in the large $N$ limit for any $m$. }
\label{fig:Bc}
\end{figure}

We plot the critical bias in Eq.~\ref{BcOddSum} as a function of $m$ and $N$ in the top panel of Figure~\ref{fig:Bc}. The bottom panel shows a cross-section of the surfaces plotted for a particular value of $m=10$. Eq.~\ref{BcEvenSum} behaves similarly as a function of $m$ and $N$. In the limit of infinite step number, where $N \gg m$, both Eqs.~\ref{BcEvenSum} and \ref{BcOddSum} converge to $|B_c| = 1/\sqrt{3} \approx 0.577$, which is apparent from Figure~\ref{fig:Bc}. Thus, any bias whose magnitude is above this critical bias $|B_c|$ will have a monotonically decreasing right tail in the PLP.

In the limit $n \gg m$, the probability of last passage (PLP) is approximately equal to the probability of last return (PLR) in the recurrence problem (i.e. returning to the origin rather than reaching a target other than the origin)~\cite{Sports,EPL}:
\begin{equation*}
\text{PLP} = p^{\frac{n-m}{2}}q^{\frac{n+m}{2}}S(N-n) \approx p^\frac{n}{2} q^{\frac{n}{2}}S(N-n) = \text{PLR}.
\end{equation*}
The PLR is typically given in the large step number (long time) limit, which is bimodal due to the functional form of its prefactor $1/\sqrt{n(N-n)}$ (e.g. see Eq.~12 of Ref.~\cite{Sports}). This large $N$ approximation features two diverging peaks of the PLR at the beginning of the walk ($n=0$) and at the end ($n=N$). Thus one might expect the same behavior for the PLP for $n \gg m$. However, our exact, closed-form results indicate that the PLP does not necessarily diverge at the end of the walk ($n=N$), in contrast to the large $N$ approximation. In particular, the critical bias condition provides a clear counterexample to the behavior predicted by the large $N$ approximation, thus underscoring the value of exact, closed-form results.

\end{document}